\documentstyle[preprint,aps]{revtex}
\tightenlines
\def\beq{\begin{eqnarray}}
\def\eeq{\end{eqnarray}}
\def\beq*{\begin{eqnarray*}}
\def\eeq*{\end{eqnarray*}}
\def\be{\begin{equation}}
\def\ee{\end{equation}}

\def\rarrow{\rightarrow}

\begin{document}
\title {\small\normalsize {\bf A NEW MEASUREMENT OF THE $^7$Li(d,p)$^8$Li\\ CROSS SECTION 
 AND CONSEQUENCES FOR $^7$Be(p,$\gamma$)$^8$B.}}
\author{ L. Weissman$^1$, C. Broude$^1$,  G. Goldring$^1$, R. Hadar$^1$,\\ 
M. Hass$^1$,  F.
Schwamm$^{1,2}$ and M. Shaanan$^3$
 \\
1.- Department of Particle Physics, The Weizmann Institute of Science \\
Rehovot, ISRAEL.\\
2.- Ruprecht-Karls Universit\"{a}et, Heidelberg, GERMANY\\
3.- Institute of Solid-State Physics, The Technion, Haifa, ISRAEL.}
\date{\today}
\maketitle
\baselineskip 18pt
\begin{abstract}
A novel scheme for measuring the cross section of the 
$^7$Be(p,$\gamma)^8$B reaction, the major source of high energy neutrinos from the sun, is presented. The scheme involves a 
strictly uniform particle beam and overcomes some of the
recognized
experimental uncertainties of previous measurements. A
new
 measurement of $\sigma[^7$Li(d,p)$^8$Li] has been carried out using this setup,  and the 
 present value
 of $\sigma[^7$Li(d,p)$^8$Li] = 155(8) mbarn at the top of the E$_d$(lab.)
= 776 keV resonance is compared to previous
measurements.  A new issue regarding both the (d,p) and (p,$\gamma$) reactions has been examined: reaction-product nuclei which are backscattered out of
the target. Measurements and simulations carried out in the course of this 
investigation are presented and discussed in the context of possible effects on the measured cross sections of these reactions.

\end{abstract}
\newpage\noindent
\section{\bf Introduction}
\par The cross section of the reaction $^7$Be(p,$\gamma)^8$B has
recently become  the subject of renewed
intense scrutiny [1] owing to the pre-eminence of the $^8$B reaction products   as the main source
of high energy neutrinos in the interior of the sun and, hence, the possible
implications for the
so-called ``solar neutrino problem". A number of new precision
measurements of this cross section  in various stages of preparation
have been reported lately. We have
set up an experiment
for such a measurement at the Weizmann Institute, focusing on one
of the major sources of
uncertainty in previous experiments:  the homogeneity of the areal
density of the target
material.
\par In general, when a nonuniform particle beam impinges on a
nonuniform target, the
reaction yield is given by: 
\be
Y=\sigma\int{dn_b\over dS}{dn_t\over dS}dS
\ee
where $n_b,n_t$ are the respective total numbers of beam and target
 particles and
${dn_b\over dS},{dn_t\over dS}$ are areal densities. \\
Only when the target is known to
be uniform and the beam
is smaller than the target can eq.(1) be simplified to:
\be
Y=\sigma {dn_t\over dS}\int{dn_b\over dS}dS = \sigma {dn_t\over dS}n_b
\ee
In such a case, the evaluation of the cross section is independent of the areal distribution of the particle beam.  On the other hand, in other cases - e.g. for radiochemically produced $^7$Be
targets [2]  - the target cannot be assumed to be 
strictly uniform and the full relation (1) has to be used in the evaluation.
 The inherent  uncertainties
in the distributions ${dn_b\over dS}$ and  ${dn_t\over dS}$ may thus
lead to considerable uncertainties in the value of the integral and hence in the deduced cross section.  We have
addressed this problem by
inverting the arrangement:  we use a homogeneous beam - produced
by raster scanning -
impinging on a target {\underline {smaller}} than the beam.
The relation (1)
then reduces to:
\be
Y=\sigma {dn_b\over dS}n_t
\ee
 independent of the potentially problematic determination of the target areal distribution.
\par As a first step we measured the cross section of the
reaction $^7$Li(d,p)$^8$Li at the top of the resonance at E$_d$(lab.)=776 keV. 
The mirror nuclei $^8$Li and $^8$B have very similar mean lives and  
 similar $\beta$ decays   
($\beta^-$  and $\beta^+$,  respectively) to $^8$Be $\rightarrow$ 2$\alpha$.  
The main differences are (i)
the much easier preparation of a $^7$Li target and (ii) the much higher yield from
$^7$Li(d,p).  The reaction
$^7$Li(d,p)$^8$Li can therefore provide a convenient check of the
equipment and the method.
Beyond that, this reaction is of significance in its own right and as a
stepping stone and
calibration in some $^7$Be(p,$\gamma$)$^8$B experiments [1-4].\\
\par
In the course of the present investigation we have examined a new issue:  reaction-product nuclei being backscattered from the target backing, leaving the target assembly and reducing the number of detected 
$\alpha$ particles. We present below experimental results which probe this issue and compare them to computer simulations. The influence of this effect on past and future measurements is discussed.

\section{\bf Experiment and Procedure}
\par The general scheme of the experiment is shown in Fig. 1a.
A d$_2$ beam
out of the Weizmann Institute 3 MV Van de Graaff accelerator is
raster scanned over a rectangle of 7 mm$\times$ 6 mm. The purpose of the scan is to obtain
a beam of uniform areal density. The d$_2$ beam (as well as the p
beam, see  below) is collimated by a 3 mm diameter hole and
impinges on a circular target of LiF of 2 mm diameter, evaporated on   Al and Pt foils (see discussion below); the target spot is 
aligned with a set of variable collimators downstream of the target. The target is mounted on an arm which is
 periodically moved out of the
beam and in front of a 40 micron surface barrier detector registering
the delayed $\alpha$'s following the $\beta$ decay of $^8Li$. 
The detector was surrounded by a  shroud to prevent 
 scattered beam particles from reaching the detector. The time sequence of the whole
cycle is: a.- 1.5 s beam-on-target; b.- 100 ms rotation; c.-  1.5 s of target in the 
counting position; d.- 100 ms rotation. In the counting position a gate signal from the control unit opens the ADC for $\alpha$
counting and the  gated scaler for 
 Faraday-cup beam monitoring. This sequence results in an efficiency factor for the
$\alpha$ count (see below) of $\eta$(cycle) = 0.400(1) (Fig. 1b).
 A  liquid nitrogen cold cryofinger is placed close to the target area to
protect the target surface from contamination. The  vacuum in the chamber was 
$~$ 8$\cdot$10$^{-7}$ torr.  

The beam density ${dn_b\over dS}$ was measured by  collimating the beam by a series of 
holes of known areas
downstream from the target position, integrating the collimated beam in an electron-suppressed Faraday cup and counting the digitized counts in a gated scaler.
The current digitizer and the scaler were checked during the experiment with a a calibrated current source.
The beam homogeneity was virtually insured by the nature of the
raster operation: a low frequency triangular $y$ scan and a high frequency triangular $x$
scan, in small, digitally controlled steps in clock-fixed time intervals.
The  beam homogeneity was checked in two direct ways: 1.- by measuring the areal density of
x-rays from a tin foil induced by the scanned proton beam in a phosphor image plate ( Fig. 2a, see Ref. 5), 
2) by repeating the measurement
 with different downstream collimators of know apertures. For a homogeneous beam
 the Faraday-cup counts
of the  integrated beam, normalized to the $\alpha$ yield, should be 
proportional to the  area of the hole (Fig. 2b).
The collimator hole areas were measured to an accuracy better than 1$\%$  by a 
microscope and by having an alpha source in front of the collimator-detector 
assembly.  The number $n_t$ of $^7$Li nuclei in the 2 mm target 
was determined in a similar measurement using the reaction
$^7$Li(p,n)$^7$Be with a proton beam of 1.985 MeV ( Refs. 6 and 7). The same downstream beam collimator of 1.5 mm diameter was used in both
measurements. The number of $^7$Be's
produced was measured by registering $\gamma$'s from the beta decay $^7$Be$\rarrow^7$Li$^*$ (478 keV) in a Ge detector at the low-background counting laboratory of the NRC-Soreq Center.  We have also taken care to adjust the plate-voltage to ensure the same area for both the p and d$_2$ beams, respectively. 

For the ratio of the two yields we get from (1):
\be
$${Y(^8Li)\over Y(^7Be)} ={\sigma(^8Li)\over\sigma(^7Be)} {n_d\over{n_p}}$$
\ee
where $n_d,n_p$ refer, respectively, to the total number of deuterons and
protons registered. 
The number of $^8$Li products, Y($^8$Li), is evaluated from the measured $\alpha$ counts:
$$n_\alpha = Y(^8Li)\eta(\alpha)$$ and Y($^7$Be) is evaluated from the measured $\gamma$ counts of the radioactive decay of $^7$Be:
$$n_\gamma = Y(^7Be)\eta(\gamma)$$
The efficiency factors $\eta(\alpha)$ and $\eta(\gamma)$ above are each products of individual and independent efficiency factors detailed below. Inserting the expressions for n$_\alpha$ and n$_\gamma$ into (4) we obtain the $^7$Li(d,p)$^8$Li cross section in terms of the $^7$Li(p,n)$^7$Be cross section and measured quantities:
\be
$$\sigma_{(d,p)} = \sigma_{(p,n)} {n_\gamma\over n_\alpha}{n_p\over n_d}{\eta(\gamma)\over\eta(\alpha)}$$
\ee

\subsection{The  $\alpha$-particle counting efficiency -  $\eta(\alpha)$}
\begin{enumerate}
\item {\it The solid angle}. 
\par
The detector solid angle was defined by a
collimator of
diameter d=8.14(3) mm  which was placed at a distance 
$\ell$ = 81.92(10) mm from the target in the counting position. The
solid angle
 is given by:  $\Omega = {1\over
16}{d^2\over\ell^2}$ = 6.16(6) 10$^{-4}$
of $4\pi$.
\item {\it The time-sequence efficiency -  $\eta$(cycle)}
\par
 This factor relates to 
the time fraction of the $\alpha$ counting out of the entire irradiation-counting cycle. From the
description  above and following Refs. 2,3,  this factor is calculated to be: $\eta$(cycle) = 0.400(1).
The time intervals of the rotation cycle were measured  by counting 
 a time-reference signal in the scaler gated by the cycle time windows ( Fig. 1b ). 
 
\item {\it The spectral efficiency.}
\par
 The $\alpha$ spectrum 
in Fig. 3.  was integrated from the point marked in the figure all the way to 
the top
energy. The cutoff entailed a loss of low energy $\alpha$'s on one hand and on the other - the
inclusion of the high energy tail of the noise spectrum. These two quantities were estimated
by rough extrapolations  to be considerably less than
1$\%$ and approximately equal and hence the integration error appears to be negligible.
The low energy tail of the $\alpha$ spectrum was also evaluated directly from the measured
$\alpha$ spectrum in Ref. 8 folded with the energy loss of the $\alpha$'s in the target, yielding
a lost fraction of 0.2$\%$.\\
\end{enumerate}

\subsection{The $\gamma$ counting efficiency - $\eta(\gamma)$}
\begin{enumerate}
\item {\it The $\gamma$ branch.} 
\par
The $\gamma$ branch of the decay
of $^7$Be to the first excited state of $^7$Li at 478 keV is
0.1052(6) [9].
\item {\it Absolute calibration of the $^7$Be activity.} 
\par
The activity of the  $^7$Be 478 keV $\gamma$-ray source produced in the (p,n) reaction on the various targets was determined by 
comparison with calibrated $^{22}$Na, $^{137}$Cs and  $^{133}$Ba 
$\gamma$ sources  at a fixed distance from a Ge detector, shielded for low-background,  at the $\gamma$-counting laboratory of the 
NRC-Soreq Center. 
\footnote{ The laboratory is designated as an International Reference 
Center for Radioactivity (IRC) by the World Health Organization.}
  The absolute intensities of the $\gamma$ sources used are known to within 2.5 - 3$\%$ each.  
The  number of 478 keV $\gamma$ rays (and hence the number of $^7$Li nuclei in the target)  is thus  determined to an accuracy of 3$\%$. 
\
%\newpage\noindent
\item {\it The $\beta$ decay of $^7$Be.}  The LiF targets were bombarded by
protons for several hours and $\gamma$'s were counted for about a day a short time later. The
number of $^7$Be's was inferred from the number of $\gamma$ counts through the
relation:
$$n_\gamma = n_{Be}{t_1\over\tau}exp{{(-t/\tau)}}\eta(1)\eta(2) $$
where $t_1$ is the $\gamma$ counting time, $t$ is the time from the
middle of the $p$ run to the middle of the counting time, $\tau$ is the $^7$Be
 mean life:
$\tau = 76.88$ d [9] and
$\eta(1)$ and $\eta(2)$ are the efficiency factors referred to above.
\end{enumerate}

\subsection{ Other quantities in eq. (5)}
\begin{enumerate}
\item {\it Current integration.}  
\par
The quantity  $n_d\over n_p$ is the
ratio  of scaler counts from a current integrator monitoring the current in
the  Faraday cup for the d$_2$ and p beams, respectively, using the same
collimator hole.
The construction of the Faraday cup, including the suppressor, ensures
a reliable  monitoring of the current. A calibrated current source was used to
check the  accuracy of the current integrator. In any case, as only the $\it {ratio}$ of the counts enters the cross-section
determination,  any  possible error from this source cancels out in first order.

\item {\it Counting statistics - $n_\gamma\over n_\alpha$.} 
\par
The statistical errors of the $\alpha$ and $\gamma$ counts were typically 
0.5$\%$ and 0.7$\%$, respectively  for all the targets used. There is in addition a 1$\%$ systematic error in the $\gamma$ counting due to uncertainties in the background subtraction procedure.

\item {\it The $^7$Li(p,n)$^7$Be cross section.}
\par
The cross section for this reaction exhibits a flat and constant region around the beam energy of E$_p$ = 1.985 MeV employed in this experiment. The cross section at this energy  was taken from Refs. 6,7, which are in very good 
agreement with 
 each other.  Each of these references quotes, separately, an absolute error of
 5$\%$. We therefore adopt the value of $\sigma[^7$Li(p,n)$^7$Be] =
269(9) mb at
E$_p$ = 1.985 MeV.\\
 Table I presents a summary of the experimental errors discussed above.
 
 Another source of error which was considered is the production of $^7$Be by $^6$Li(d,n)$^7$Be,
adding on to the $^7$Be produced by $^7$Li(p,n)$^7$Be. The added fraction was evaluated from the
 known cross-sections [10] and the fraction of $^6$Li in the target to be less than 0.2$\%$
and therefore insignificant.\\
A larger correction to the extracted cross section can arise
 from the effect of $^8$Li lost from the target due to backscattering. 
This issue is discussed  in III.  
\end{enumerate}

\subsection{d$_2$ and p energy calibration.} 
A d$_2$ beam at the 776 keV
resonance corresponds to an 
 accelerator voltage of 1.552 MV. The voltage was calibrated using
the known  thresholds of the $^{13}$C(p, $\gamma$ = 9.17 MeV)$^{14}$N
 reaction at 1.746 MeV and the $^7$Li(p,n)$^7$Be reaction at 1.88 MeV which are close to the d$_2$ and p beam voltages. 

\vskip .5cm\noindent

\section{Backscattering of reaction-product nuclei}
In the course of this experiment it became apparent that under some conditions
the backscattering  of recoil nuclei out of the target can  be significant and 
can affect the cross section measurements. In order to obtain a quantitative 
estimate of these effects, the propagation of the 
 reaction products of the $^7$Li(d,p)$^8$Li and $^7$Be(p,$\gamma$)$^8$B 
reactions was simulated with the aid of the TRIM [11] code. We have also determined these effects experimentally for the (d,p) reaction.

\subsection{Backscattering for $^7$Li(d,p)$^8$Li and $^7$Li(p,n)$^7$Be}
For the $^7$Li(d,p)$^8$Li reaction one has to take into account the broad 
angular distribution of the protons [12] (and  
the $^8$Li recoils ) and also to evaluate the loss of the backscattered $^7$Be's
from $^7$Li(p,n)$^7$Be reaction, used as a standard (see section II).  One can 
then infer the total effect of backscattering loss on the cross-section measurement.

 Calculations were carried out for
 LiF targets on Pt, Ni and Al backings at 776 keV deuteron energy. The results presented here are an average of simulations of reactions occuring at 
three positions in the LiF target. At a given depth in the LiF target the 
energy and angular distribution of $^8$Li ions were evaluated  and used to obtain the angle and energy distributions of the backscattered $^8$Li ions from TRIM.  The backscattered nuclei were then traced in the direction opposite 
to the beam to obtain the number of ions leaving the target. 

A similar procedure was carried out for
the backscattered $^7$Be from the $^7$Li(p,n)$^7$Be reaction for the proton beam energy of 1.99 MeV. The angle and energy distributions of
the neutrons (and the $^7$Be recoils) were taken from Ref 13. The results of the TRIM simulation
are shown in Table II. As we see from the table, the effect of backscattered particles can give rise to a sizeable  correction for high Z substrates.
 
\subsection{backscattering for $^7$Be(p,$\gamma$)$^8$B}
  Similar backscattering effects also  exist for the $^7$Be(p,$\gamma$)$^8$B reaction. In this case, the kinematics of the reaction is much 
simplified as the $^8$B nuclei from the proton-capture reaction all proceed in the proton-beam direction with the same center-of-mass momentum. As an example we
have performed calculations which duplicate the conditions of Ref. 2. In that experiment 
the target was electroplated on polished  Pt disks and  contained 7 $\mu$g 
of solid material with 11$\%$ of $^7$BeO [14]. For low proton energy, the number of 
backscattered particles depends very strongly on the (unknown) target 
composition. Two hypothetical trial cases are 
shown in Fig. 4. The first is for  $^7$BeO dissolved in pure iron; the second case
is for a hypothetical composition of equal number of CuO,  Cu$_2$N$_3$ and C molecules. The latter
composition  has been chosen
following  the statement in Ref. 14 that the solid materials remaining after the plating and flaming procedure 
are most likely to be traces of light metal oxides, nitrates and carbon. 
It is obvious from Fig. 4 that heavy atoms in the target and target backing should be avoided as much as possible in precision measurements of the 
$^7$B(p,$\gamma$)$^8$B cross section.

\subsection{Backscattering measurements for the $^7$Li(d,p)$^8$Li reaction.}

We have used several LiF targets with  Al and Pt backing materials in order to
probe experimentally 
 the backscattering issue. The results
of the cross section measurements for five targets, together with the backscattering fraction as calculated by TRIM,   
 are presented in Table III. The error of each measurement is about 5$\%$
( Table I ). The experiments yield
 a 9.1(2.2) $\%$ difference between the 400 $\AA$  LiF on Pt and the LiF on Al targets.
The relative error for the two types of targets is smaller than the individual  errors since it does not include
the common uncertainties of the $^7$Li(p,n)$^7$Be cross section and the 
$\gamma$-detector 
efficiency. The TRIM simulations are in good agreement with
the experimental results.

\section{Results and Conclusion.}
\vskip .5cm\noindent
As the  final result of the $^7$Li(d,p)$^8$Li cross section
we have adopted  an average of the two aluminum-backed
 targets, believed to be free from the backscattering loss. 
We get: $\sigma$[$^7$Li(d,p)$^8$Li] = 155(8) mbarn. \\
A recommended value for this cross section is given in Ref. 1 : 
$\sigma$ = 147(11) mbarn. This value is based on four measurements, two of which should be corrected for backscattering loss. Based on the available target details as presented in the respective publications, we estimate the corrected value to be in the range 150 - 152 mbarn, very close to our value. 
\par
Previous measurements of the $\sigma(^7$Be((p,$\gamma)^8$B cross section are also susceptible 
to backscattering losses.  In particular, the results of Ref. 2 should be corrected for both the 
$^7$Be(p,$\gamma$)$^8$B and the  $^7$Li(d,p)$^8$Li reactions. 
The $^7$Be(p,$\gamma$)$^8$B reaction was measured at various energies
and, as can be seen from Fig. 4, the backscattering correction diminishes at low energies.
The $^7$Li(d,p)$^8$Li reaction at the resonance energy was then used by the authors  
 to obtain the number of $^7$Li, and hence $^7$Be, atoms in the same platinum-backed target. 
Since the correction  due to this is a $\it constant$ which is of the opposite sign and larger than the $^8$B corrections ( Table II and Fig. 4),  the overall effect is to reduce the derived S factor, especially at lower energies.
\par

 We would like to express our sincere thanks to Ygal Shahar for his expert help in operating 
the Van de Graaff accelerator and in constructing the beam line. We wish to thank  Meir Birk  and Moshe Sidi for designing and constructing the raster scanner, and Leo Sapir and Boris Levine
for the target preparation. We acknowledge with thanks the expert help of Ovadya Even of the Soreq Center in the $\gamma$-efficiency calibration. We are grateful to Prof. Claus Rolfs for extensive and enlightening 
 discussions.

\vskip 2cm\noindent

\begin{figure}
\caption{1a - Schematic view of beam line; the beam scanner and
measuring chamber are indicated. 1b - The time sequence of the  
stepping motor rotation, irradiation and data acquisition.}
\end{figure}

\begin{figure}
\caption{Scanned-beam uniformity measurements.  a - A one-dimensional intensity cut of the x-ray yield from the molecular plate [5]. b - The d$_2$ count in the Faraday cup, normalized to
the $\alpha$ yield, as a function of collimator area.}
\end{figure}

\begin{figure}
\caption{A typical $\alpha$ spectrum.}
\end{figure}

\begin{figure}
\caption{
Results of the TRIM simulations for the $^8$B backscattering loss  
for the conditions of Ref. 2, assuming two different target compositions: 
 circles - $^7$BeO  dissolved in pure iron;  squares - $^7$BeO dissolved in equal amounts of CuO, Cu$_2$N$_3$ and C molecules.} 
\end{figure}

\begin{table}
\caption{ Compilation of experimental errors.}

\baselineskip=12pt
\begin{tabular}{cc}
\hline
\hline
The source of the experimental error& Error ($\%$)\\
\hline
\hline
 Uncertainty in $\sigma_{(p,n)}$&  3.5 \\
 Absolute efficiency of $\gamma$ counting& 3 \\

 Systematic error in $\gamma$ counting& 1\\
 Statistics of gamma counting& 0.7\\
 Statistics of alpha counting& 0.5\\
 Solid angle of $\alpha$ detector& 0.5 \\
 $\gamma$ branch for $^7$Be decay& 0.6\\
 Timing of the rotating cycle&0.3\\
\hline
 Total & $\approx$ 5\\
\end{tabular}
\end{table}

\begin{table}
\caption{ The results of TRIM simulation for the number of backscattered particles
in $^7Li$(d,p)$^8$Li and $^7$Li(p,n)$^7$Be reactions for three different backing materials}
\baselineskip=12pt
\begin{tabular}{ccccc}
\hline
\hline
Backing &  Thickness of LiF&   Lost fraction& Lost fraction & Correction\\
material & (\AA) & of $^8$Li ($\%$) & of $^7$Be ($\%$) & to cross-section ($\%$)\\ 
\hline
\hline
Pt& 1000& 12.9(1.5) & 5.1(1.0) & 7.4(1.8)\\
Pt& 400& 15.6(1.5) & 5.8(1.0) & 9.8(1.8)\\
Ni& 1000& 2.4(1) & 1.3(0.7) & 1.1(0.9)\\
Al& 1000& 0.3(0.5) & 0.07(0.1) &0.25(0.6)\\

\end{tabular}
\end{table}

\begin{table}
\caption{ The results of cross section measurements for the different targets. }
\baselineskip=12pt
\begin{tabular}{ccccc}
\hline
\hline
Backing material&  Thickness of LiF & $\sigma_{d,p}/\sigma_{p,n}$ & $\sigma_{d,p}$ (mbarn) & $\sigma_{d,p}$ (mbarn) \\
&\AA&&(mbarn)& corrected by TRIM \\
\hline
\hline
Al& 1100 & 0.57(2)& 154(8)&154(8)\\
Al& 1100 & 0.57(2)& 156(8)&156(8)\\
Pt& 1200 & 0.54(2)& 146(8)&156(10)\\
Pt& 400 & 0.53(2)& 143(8)&157(10)\\
Pt& 400 & 0.52(2)& 140(8)& 154(10)\\
\hline
Adopted value&&& 155(8)

\end{tabular}
\end{table}

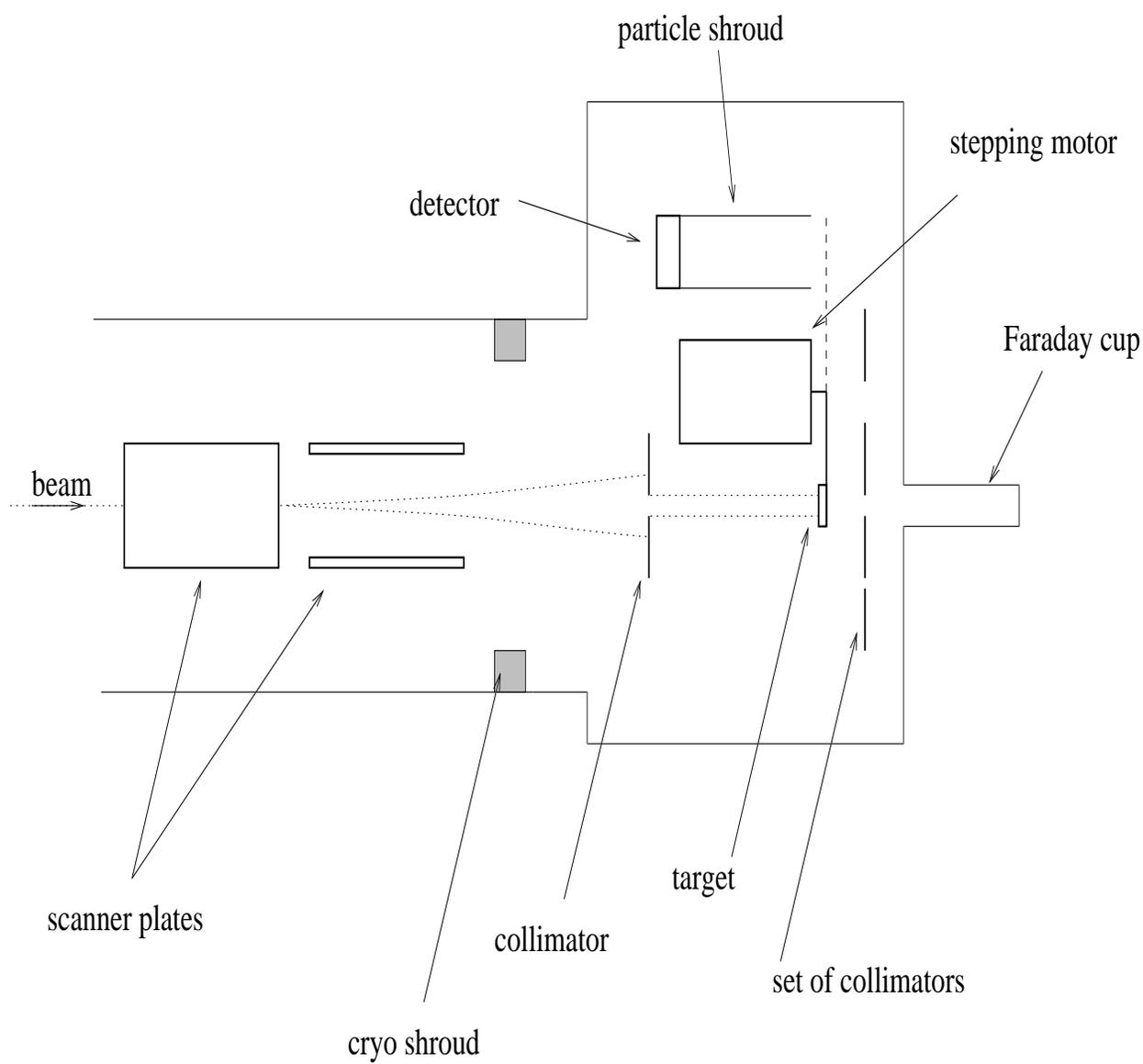
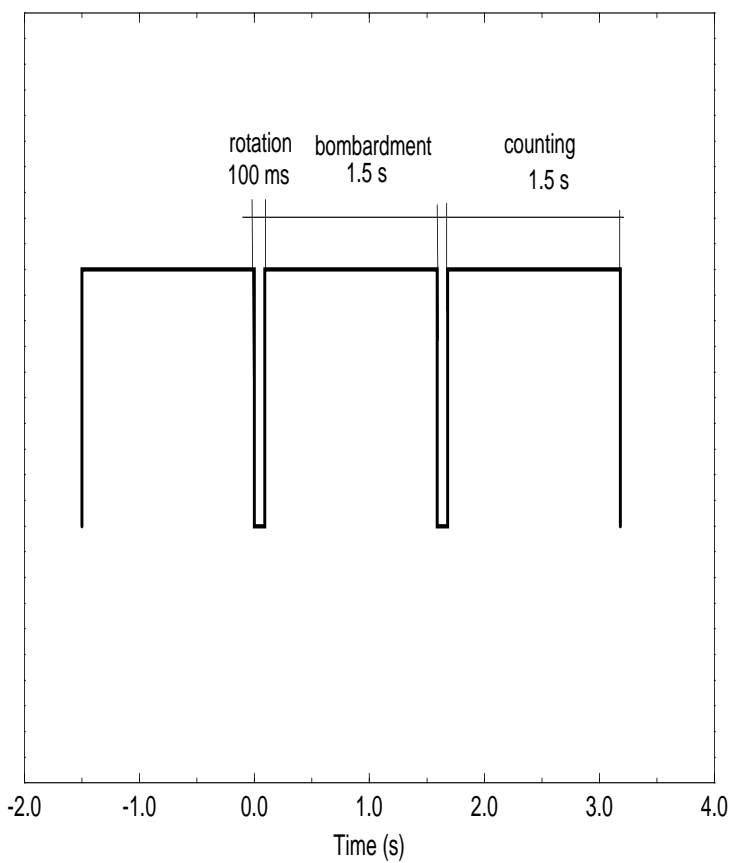
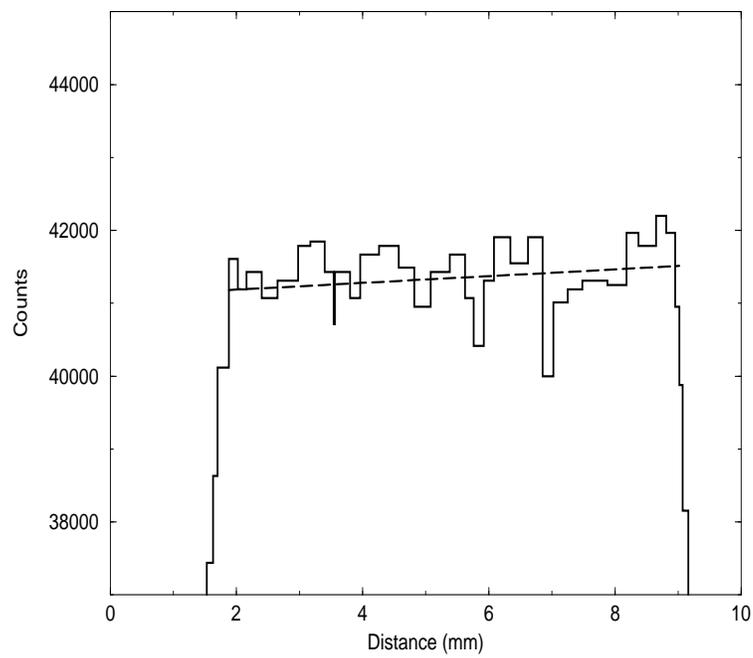
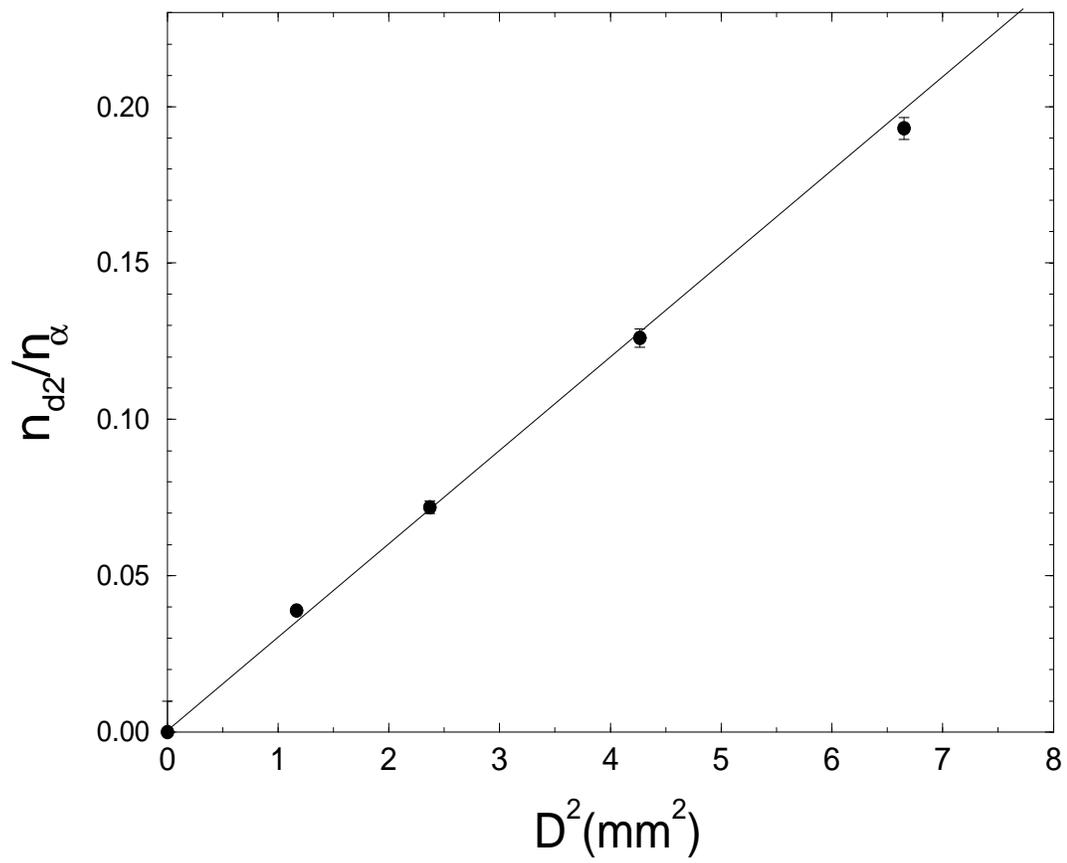
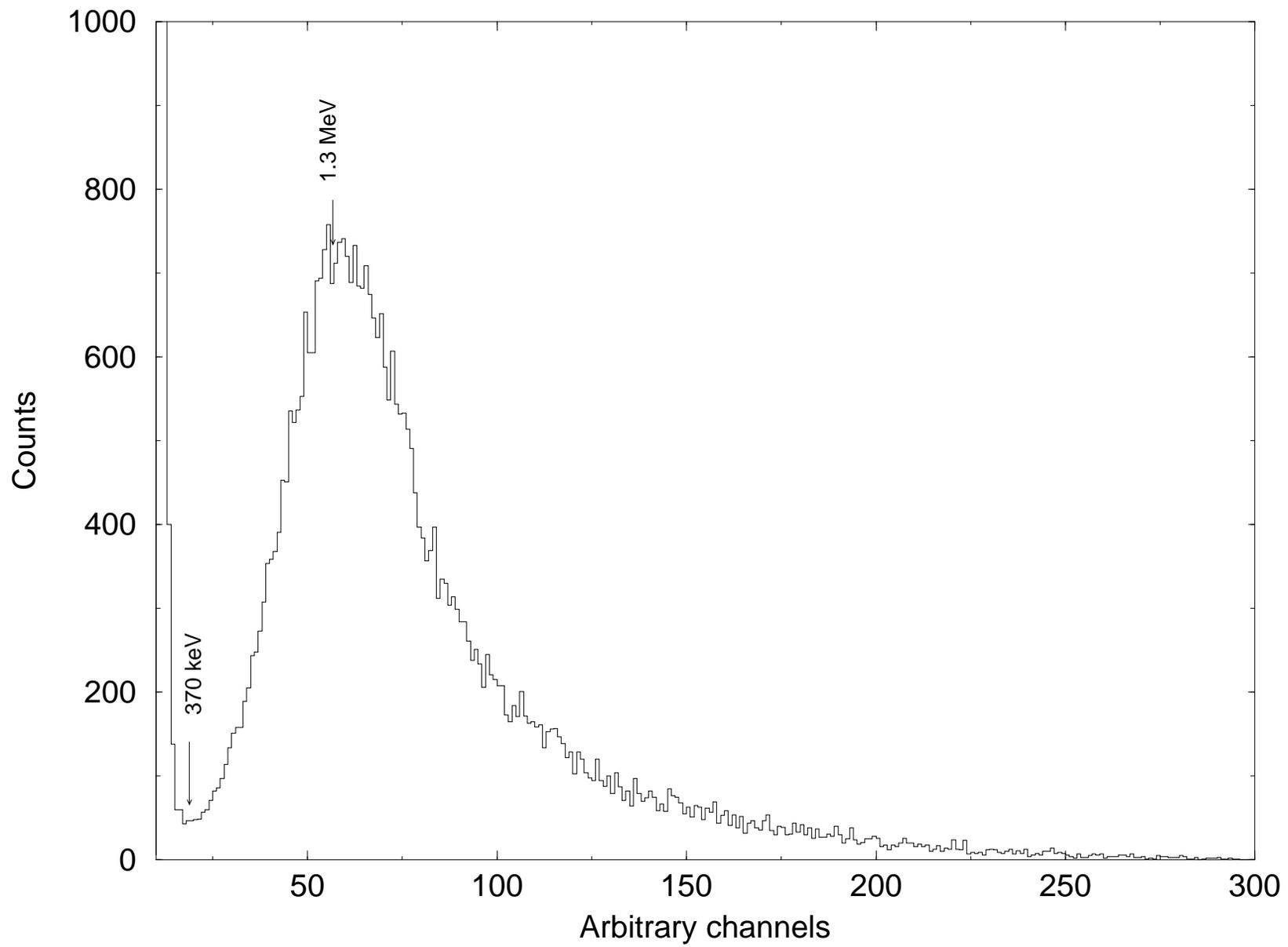
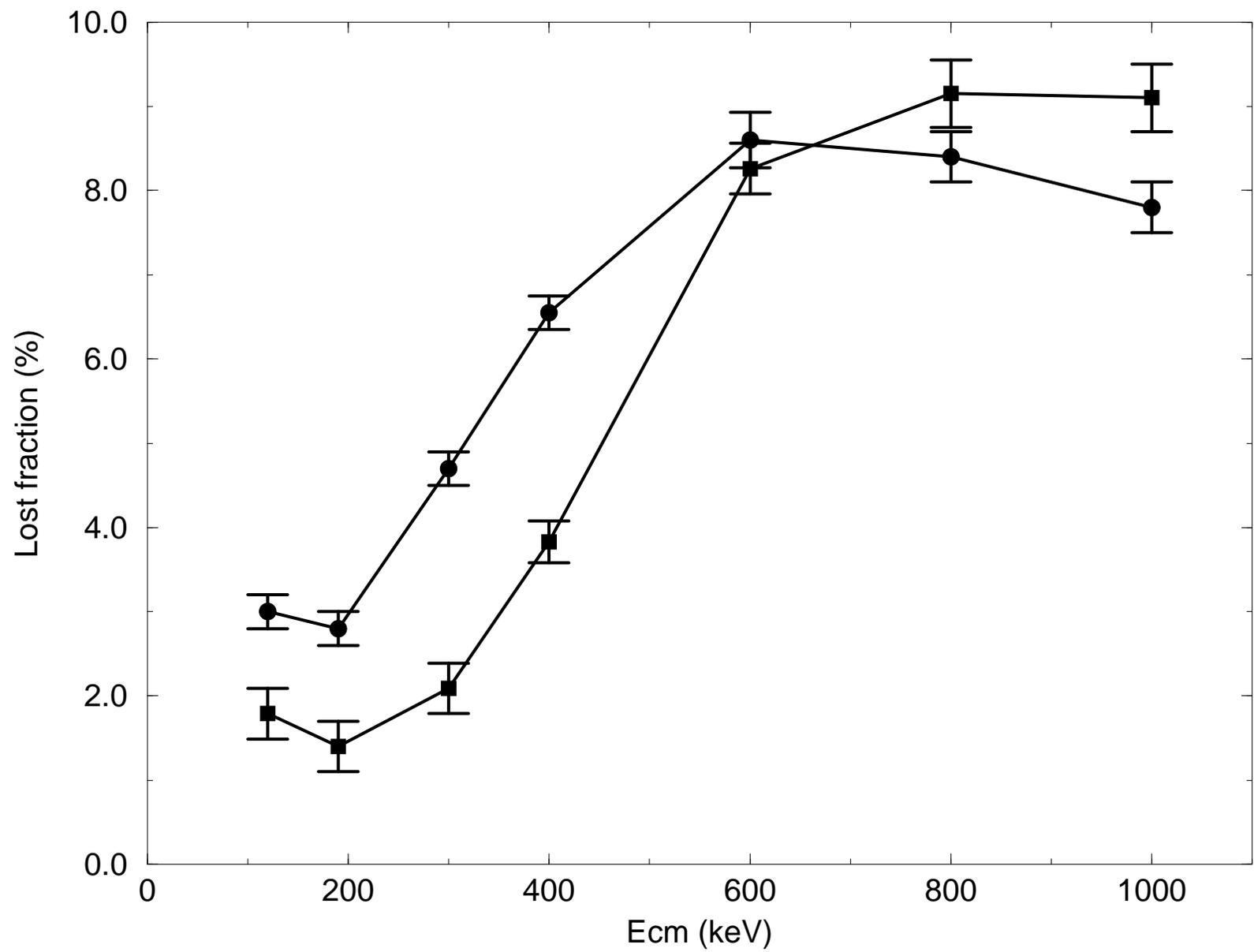
\end{document}